\documentclass[twoside,twocolumn,9pt]{article}
\usepackage{extsizes}
\usepackage[super,sort&compress,comma]{natbib} 
\usepackage[version=3]{mhchem}
\usepackage[left=1.5cm, right=1.5cm, top=1.785cm, bottom=2.0cm]{geometry}
\usepackage{balance}
\usepackage{mathptmx}
\usepackage{sectsty}
\usepackage{graphicx} 
\usepackage{lastpage}
\usepackage[format=plain,justification=justified,singlelinecheck=false,font={stretch=1.125,small,sf},labelfont=bf,labelsep=space]{caption}
\usepackage{float}
\usepackage{fancyhdr}
\usepackage{fnpos}
\usepackage[english]{babel}
\addto{\captionsenglish}{%
  \renewcommand{\refname}{Notes and references}
}
\usepackage{array}
\usepackage{droidsans}
\usepackage{charter}
\usepackage[T1]{fontenc}
\usepackage[usenames,dvipsnames]{xcolor}
\usepackage{setspace}
\usepackage[compact]{titlesec}
\usepackage{hyperref}
\usepackage{epstopdf}%This line makes .eps figures into .pdf - please comment out if not required.

\definecolor{cream}{RGB}{222,217,201}

\begin{document}

\pagestyle{fancy}
\thispagestyle{plain}
\fancypagestyle{plain}{
%%%HEADER%%%
\renewcommand{\headrulewidth}{0pt}
}
%%%END OF HEADER%%%

%%%PAGE SETUP - Please do not change any commands within this section%%%
\makeFNbottom
\makeatletter
\renewcommand\LARGE{\@setfontsize\LARGE{15pt}{17}}
\renewcommand\Large{\@setfontsize\Large{12pt}{14}}
\renewcommand\large{\@setfontsize\large{10pt}{12}}
\renewcommand\footnotesize{\@setfontsize\footnotesize{7pt}{10}}
\makeatother

\renewcommand{\thefootnote}{\fnsymbol{footnote}}
\renewcommand\footnoterule{\vspace*{1pt}% 
\color{cream}\hrule width 3.5in height 0.4pt \color{black}\vspace*{5pt}} 
\setcounter{secnumdepth}{5}

\makeatletter 
\renewcommand\@biblabel[1]{#1}            
\renewcommand\@makefntext[1]% 
{\noindent\makebox[0pt][r]{\@thefnmark\,}#1}
\makeatother 
\renewcommand{\figurename}{\small{Fig.}~}
\sectionfont{\sffamily\Large}
\subsectionfont{\normalsize}
\subsubsectionfont{\bf}
\setstretch{1.125} %In particular, please do not alter this line.
\setlength{\skip\footins}{0.8cm}
\setlength{\footnotesep}{0.25cm}
\setlength{\jot}{10pt}
\titlespacing*{\section}{0pt}{4pt}{4pt}
\titlespacing*{\subsection}{0pt}{15pt}{1pt}
%%%END OF PAGE SETUP%%%

%%%FOOTER%%%
\fancyfoot{}
\fancyfoot[LO,RE]{\vspace{-7.1pt}\includegraphics[height=9pt]{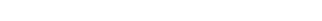}}
\fancyfoot[CO]{\vspace{-7.1pt}\hspace{13.2cm}\includegraphics{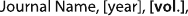}}
\fancyfoot[CE]{\vspace{-7.2pt}\hspace{-14.2cm}\includegraphics{head_foot/RF}}
\fancyfoot[RO]{\footnotesize{\sffamily{1--\pageref{LastPage} ~\textbar  \hspace{2pt}\thepage}}}
\fancyfoot[LE]{\footnotesize{\sffamily{\thepage~\textbar\hspace{3.45cm} 1--\pageref{LastPage}}}}
\fancyhead{}
\renewcommand{\headrulewidth}{0pt} 
\renewcommand{\footrulewidth}{0pt}
\setlength{\arrayrulewidth}{1pt}
\setlength{\columnsep}{6.5mm}
\setlength\bibsep{1pt}
%%%END OF FOOTER%%%

%%%FIGURE SETUP - please do not change any commands within this section%%%
\makeatletter 
\newlength{\figrulesep} 
\setlength{\figrulesep}{0.5\textfloatsep} 

\newcommand{\topfigrule}{\vspace*{-1pt}% 
\noindent{\color{cream}\rule[-\figrulesep]{\columnwidth}{1.5pt}} }

\newcommand{\botfigrule}{\vspace*{-2pt}% 
\noindent{\color{cream}\rule[\figrulesep]{\columnwidth}{1.5pt}} }

\newcommand{\dblfigrule}{\vspace*{-1pt}% 
\noindent{\color{cream}\rule[-\figrulesep]{\textwidth}{1.5pt}} }

\makeatother
%%%END OF FIGURE SETUP%%%

%%%TITLE, AUTHORS AND ABSTRACT%%%
\twocolumn[
  \begin{@twocolumnfalse}
{\includegraphics[height=30pt]{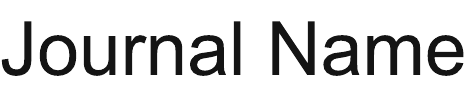}\hfill\raisebox{0pt}[0pt][0pt]{\includegraphics[height=55pt]{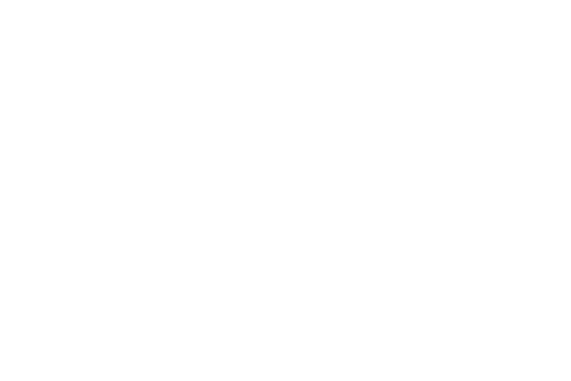}}\\[1ex]
\includegraphics[width=18.5cm]{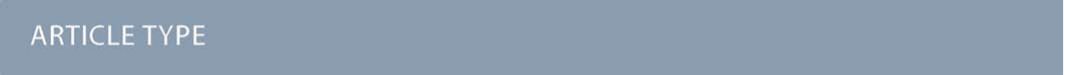}}\par
\vspace{1em}
\sffamily
\begin{tabular}{m{4.5cm} p{13.5cm} }

\includegraphics{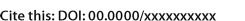} & \noindent\LARGE{\textbf{Cooperation, competition and multi-scale reorganisation: reconfigurable assembly of spherical colloids under combined electric and magnetic fields$^\dag$}} \\%Article title goes here instead of the text "This is the title"
\vspace{0.3cm} & \vspace{0.3cm} \\

 & \noindent\large{Indira Barros,\textit{$^{a}$} Sayanth Ramachandran,\textit{$^{a,b}$} and Indrani Chakraborty$^{\ast}$\textit{$^{a}$}} \\%Author names go here instead of "Full name", etc.

\includegraphics{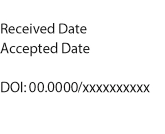} & \noindent\normalsize{
Field-induced assembly of complex, reconfigurable structures has recently gained significant attention for its potential in programmable self-assembly and micro/nanofabrication. While electric and magnetic fields can independently form diverse colloidal structures, combining them offers enhanced control, design flexibility and reconfigurability. Here we demonstrate multi-scale reorganisation of colloidal superstructures assembled under combined AC electric and DC magnetic fields, where comparable dipolar and electrohydrodynamic interactions create a rich, tunable phase space. Unlike prior studies which were mostly limited to purely dipolar regimes, our results reveal cooperative, competitive, and multi-scale, multi-stage reorganisation modes depending on field parameters. We explicitly demonstrate reversible switching between a wide range of structural configurations. Imaging and Fourier analysis show how field strength, frequency, direction, and sequence dictate structure formation. We further demonstrate that introducing an additional level of complexity through binary particle systems yields unexpected structures. This versatile, multi-field strategy enables bottom-up fabrication of adaptive, functional microsystems.} \\%The abstrast goes here instead of the text "The abstract should be..."

\end{tabular}

 \end{@twocolumnfalse} \vspace{0.6cm}

  ]
%%%END OF TITLE, AUTHORS AND ABSTRACT%%%

%%%FONT SETUP - please do not change any commands within this section
\renewcommand*\rmdefault{bch}\normalfont\upshape
\rmfamily
\section*{}
\vspace{-1cm}

%%%FOOTNOTES%%%

\footnotetext{\textit{$^{a}$~Birla Institute of Technology and Science K.K. Birla Goa Campus, NH17B Bypass road, Sancoale, Goa, India. Tel: +91 9421281860; E-mail: p20210065@goa.bits-pilani.ac.in}}
\footnotetext{\textit{$^{b}$~Max-Planck-Institut for Polymer research Mainz, Ackermannweg 10, 55128 Mainz, Germany. }}

%Please use \dag to cite the ESI in the main text of the article.
%If you article does not have ESI please remove the the \dag symbol from the title and the footnotetext below.
\footnotetext{\dag~Supplementary Information available: [details of any supplementary information available should be included here]. See DOI: 00.0000/00000000.}
%additional addresses can be cited as above using the lower-case letters, c, d, e... If all authors are from the same address, no letter is required

%\footnotetext{\ddag~Additional footnotes to the title and authors can be included \textit{e.g.}\ `Present address:' or `These authors contributed equally to this work' as above using the symbols: \ddag, \textsection, and \P. Please place the appropriate symbol next to the author's name and include a \texttt{\textbackslash footnotetext} entry in the the correct place in the list.}

%%%END OF FOOTNOTES%%%

\section{Introduction}
Multi-level or hierarchical structures emerge when different local interactions compete, cooperate or contribute across multiple length scales, leading to complex but adaptable organisation. This kind of multi-scale self-assembly is common in nature, where local interactions give rise to larger emergent structures. A good natural example is found in fire ant rafts, where thousands of ants collectively form a floating structure during floods. At a smaller scale, ants link together into clusters, while at a larger scale, the raft spreads out, balancing attractive and extensile forces \cite{mlot2011fire}. The cytoskeleton of living cells is another example that illustrates how competing interactions create adaptive structures: microtubules are long, stiff filaments that provide long-range cargo transport and resist compression, while actin filaments form dense, branched, flexible networks that shape the cell at smaller scales.\cite{15vm-4d7b} The mechanical competition between them enables the cell to switch between stable, contractile states and dynamic, exploratory ones. Inspired by such biological systems, recent advances have demonstrated that hierarchical self-assembly pathways can be encoded and programmed in synthetic colloidal and molecular systems \cite{morphew_dwaipayan2015hierarchical, morphew_dwaipayan2018programming, rao_dwaipayan2020leveraging, hayes2021encoding}. By drawing inspiration from such natural systems, functional materials can be designed with unprecedented versatility, adaptability and controllability for desired applications.\cite{MartinezACSNano,WeiResearch} From micro-robotics to targeted drug-delivery to the design of smart materials and switchable photonic crystals, programmable self-assembly of simple building blocks into complex, hierarchical structures finds enormous application potential.

One of the simplest, most widely used and inexpensive methods for obtaining adaptable, complex organisation of microparticles is field-based assembly using electric, magnetic or optical fields\cite{gong2002two,pan2024activenanorods, shah2012liquid}. Even with simple building blocks like spherical colloids with isotropic interactions, alternating current (AC) electric fields have been used to produce colloidal crystals, \cite{Trau1996_ACAssembly,lumsdon2004two} colloidal molecules,\cite{MaWu_oligo_2013} linear chains of tunable stiffness,\cite{Vutukuri_tunable} bands composed of zigzag patterns,\cite{KatzmeierPRL2022_patterns} glassy clusters and domains,\cite{barros2025assembly} and open colloidal lattices.\cite{Jingjing_Wu_openlattice} A complex interplay of the frequency and amplitude of the applied AC electric field, as well as the salt concentration of the solvent and zeta potential of the particles, determines the final structural configuration. On the other hand, spherical superparamagnetic particles under purely magnetic fields have been observed to produce linear chains and staggered chains \cite{smallenbarrosurg2012self}. The application of a bi-axial field of similar type or two different types of fields applied along different directions adds one more level of complexity to the growth of colloidal superstructures\cite{leunissen2009directing, camacho2025template}. 

A higher degree of controllability and precision in structure formation can be achieved by a combination of magnetic and electric fields. For magnetic field based assembly, the interaction is dipolar as there are no magnetic monopoles. On the other hand, electric fields offer a much richer phase space where, in addition to dipolar and screened coulombic interactions, the presence of ions in the solvent and a conducting electrode surface leads to electrohydrodynamic flows (EHD) and phenomena like induced charge electro-osmosis (ICEO) and concentration polarisation electro-osmosis \cite{dobnikar2013emergent, diwakar2022ac, katzmeier2023microrobots}. The EHD flow near the surface of an electrode can break the symmetry in this case, leading to the propulsion of even spherical particles across the electrode surface\cite{barros2025assembly} similar to an active particle. A combination of electric and magnetic fields applied along specific directions can therefore be used to create an immensely adaptable and controllable structure, while integrating additional functionalities like directed propulsion as a micro-swimmer. A summary of the types of structures or entities formed from a combination of electric and magnetic fields is given in Table \ref{table1}. Haque \textit{et al.} produced a microscopic robot by assembling a colloidal chain using an applied magnetic field and generating propulsion with an applied out of plane AC electric field.\cite{haque2023propulsion} Bharti \textit{et al.} showed that by applying an electric and a magnetic field in perpendicular directions, bidirectional particle chains, colloidal networks, and discrete crystals can be formed.\cite{bharti2016multidirectional} Demirors \textit{et al.} produced an interdigitated design of micromagnet and microfabricated electrode geometries that allowed simultaneous spatial control of the colloids by magnetophoresis and dielectrophoresis with very fast switching times.\cite{demirors2016periodically, demirors2017colloidal} Zhu \textit{et al.} demonstrated the propulsion of asymmetric magnetic dimers under orthogonally applied electric and magnetic fields.\cite{zhu2021synthesis} Assembly of these asymmetric dimers into homochiral clusters was obtained by Zhu \textit{et al.} using crossed electric and magnetic fields, where the chirality was controlled by adjusting the direction and strength of the magnetic field and the frequency of the electric field \cite{zhu2025reconfigurable, zhang2025numerical}. Recently, Haque \textit{et al.} obtained highly aligned dense but separated linear chains at intermediate particle concentrations. At higher particle concentrations, a hierarchical structure was obtained where individual particles comprising the chain assembled into colloidal oligomers of different valences \cite{haque2025high}. Demirörs \textit{et al.} have shown active cargo transport with Janus colloidal shuttles under combined electric and magnetic fields \cite{demirors2018active}, while Alapan \textit{et al.} used an out of plane AC electric field to drive magnetic microparticles into the wheel pockets of a microcar and propelled it using a rotating magnetic field.\cite{alapan2019shape} Despite the extensive body of existing research on single-field-induced assembly and propulsion of colloids \cite{liljestrom2019active,harraq2022field}, work on multi-field induced assembly of reconfigurable colloidal superstructures are relatively less. Additionally, in multi-field assembly strategies, mostly dipolar interactions have been utilized till date to produce complex hierarchical structures.\cite{bharti2016multidirectional, haque2025high} On the other hand, to generate propulsion, dielectrophoretic and magnetophoretic forces have been utilized.\cite{demirors2016periodically,demirors2017colloidal,demirors2018active} In the vast majority of studies on multi-field induced assembly, the discussion involves assembly under dipolar forces, and EHD forces either appear as an additional consideration or are not of primary importance at all. In contrast, our work explicitly employs the EHD forces and their variation at different frequency regimes to design reconfigurable structures with a wide range of structural configurations. We show that in systems where electrohydrodynamic forces are non-negligible, a competition between the dipolar and EHD forces lead to fundamentally different structure formation and phase behaviours.

In this work, we demonstrate field-based assembly of complex hierarchical structures in a frequency and particle size regime where magnetic dipolar interactions and electric dipolar and electrohydrodynamic interactions are comparable, so they can lead to three distinct behaviours--a) cooperation, b) competition and c) multi-scale reorganisation. Magnetic dipolar interactions tend to drive particles into linear chains or vertical columnar stacks, depending on the field direction. In contrast, electric dipolar interactions combined with EHD flows from an out-of-plane AC electric field can generate either stacked columns or extended quasi-planar, glassy structures at larger length scales, depending on the frequency. The subtle interplay between the electric and magnetic fields leads to reinforced stacking (cooperation) or disintegration of larger stacks into smaller stacks (competition). Additionally, for crossed electric and magnetic fields, this leads to the formation of a two-level structure where each micrometre-sized colloidal particle (`building block’) organises into a linear arrangement over smaller length scales, whereas groups of such linear chains arrange into clustered `domains’ over a length scale of tens of micrometres. An explicit demonstration of the reconfigurability of our system highlights the significant application potential of our system in bottom-up assembly of reconfigurable and smart materials \cite{morales2016bending}, switchable photonic crystals \cite{liu2023reconfigurable, liao2021colloidal}, and even in the fabrication of modular micro-swimmers for targeted drug-delivery \cite{zhu2022external, wang2023multimode} and environmental remediation.

\begin{table*}
\small
\caption{\ Experimentally realised structures and entities from combined electric and magnetic fields}
\label{table1}

\begin{tabular*}{\textwidth}{@{\extracolsep{\fill}}
p{1.7cm} % Paper
p{2.2cm} % Particle details
p{1.5cm} % Fields
p{3.0cm} % Field parameters
p{2.3cm} % Final structure
p{3.0cm} % Forces
}
\hline
Paper & Particle details & Field(s) & Field parameters & Final structure & Forces responsible \\
\hline
\\

Haque \textit{et al.}\cite{haque2023propulsion} &
1 $\mu m$ Dynabeads, magnetic &
$E_z$, $B_y$, $B_z$ &
$E_z$ 0.5–50 kHz, 2–20 Vpp;  
$B_y$ (chain formation), $B_z$ (tilting) = 0–16 mT, 20-60 Hz &
Translating chains &
Magnetic dipolar forces form and tilt chains; EHD gives propulsion from tilted asymmetry \\
\\

Bharti \textit{et al.}\cite{bharti2016multidirectional} &
5.7 $\mu m$ Superparamagnetic PS &
$E_{y}$, $B_{x}$ &
$E_{y}$=10 kHz, 10–20 V/mm;  $H_x$=60-125 A/m &
Grid-like networks (E $\perp$ B), elongated 2D hcp crystals (E and B at 45$^\circ$) &
Electric and magnetic dipolar forces \\
\\

Demirors \textit{et al.}\cite{demirors2016periodically} &
8.3 $\mu m$ SiO$_2$ coated alumina platelets containing superparamagnetic iron oxide nanoparticles &
$E_z$, $B_{3D, rotating}$ &
$E_z$ =1 MHz, 1 $V_{rms}/\mu m$; $B$ = 0.4 T, 1.5 Hz &
Ordered microstructures of tunable shape and mechanical properties &
Dielectrophoretic forces control position, magnetic dipolar forces control orientation of particles \\
\\

Demirors \textit{et al.}\cite{demirors2017colloidal} &
4.69 $\mu m$ and 2.85 $\mu m$ super-paramagnetic polystyrene &
$E_z$, in-plane B &
$E_z$=1 MHz, 0.25 V/$\mu m$; $B_{planar}$=2–3 mT (inhomogenous) &
Rapid switching  between magnetic and electric potential minima &
Dielectrophoresis (DEP), magnetophoresis \\
\\

Zhu \textit{et al.}\cite{zhu2021synthesis} &
3 $\mu m$ polystyrene dimers coated with Fe$_3$O$_4$ nanoparticles and SiO$_2$ shell &
$E_z$, $B_{planar} (rotating)$ &
$E_z$ =0.4–1 kHz, 4- 20 Vpp;
$B_{planar}$ =0.83–5.22 mT, 0-50Hz &
Active particle &
EHD propels dimers, B steers.\\
\\

Zhu \textit{et al.}\cite{zhu2025reconfigurable} &
2-6 $\mu m$ polystyrene dimers coated with Fe$_3$O$_4$ nanoparticles+ SiO$_2$ shell &
$E_z$, $B_{planar} (rotating)$ &
$E_z$=0.5-5 kHz, $10^4$-$10^5$ V/m; $B_{xy, rot}$=0.4–1.06 mT, 50 Hz &
3D chiral/achiral structures, 2D clusters &
Electric dipolar and EHD interactions form chiral clusters, magnetic dipolar forces selects chirality \\
\\

Haque \textit{et al.}\cite{haque2025high} &
1.05 $\mu m$ Dynabeads (carboxyl functionalized), magnetic &
$E_z$, $B_{planar}$ &
$E_z$ = 350 Hz-1 MHz, 0-10 Vpp; $B_y$=4-5 mT, 50 Hz&
Oligomers, high density oligomer chain arrays &
Primarily electric and magnetic dipolar, EHD facilitates oligomer formation.\\
\\

Demirors \textit{et al.}\cite{demirors2018active} &
4.6 $\mu m$ SiO$_2$ coated with Ni + Pt (Janus) &
$E_z$, $B_{planar}$ &
$E_z$=0.5–2.5 MHz, 0.03–0.06 V/$\mu m$;  
$B_{xy}$=0.2 T &
Janus shuttles &
Self-DEP propulsion, dipole–dipole cargo binding, magnetic steering \\
\\

Alapan \textit{et al.}\cite{alapan2019shape} &
10 $\mu m$ superparamagnetic polystyrene &
$E_z$, $B_{x,y,z}$ &
$E_z$ = 4 kHz-1 MHz, 6 Vpp; $B_x,y,z$=10-20 mT&
Wheels of a microcar &
DEP for assembly, magnetic torque for wheel rotation (propulsion) and steering\\
\\

Han \textit{et al.}\cite{HanPNAS2017} &
3 $\mu m$ SiO$_2$ coated with Ni, Ti (Janus) &
$E_z$, $B_{planar}(rotating)$ &
$E_z$ = 5 kHz, 7 V square wave; $B_{xy, rot}$=5 mT, 0.05-16 Hz&
Two groups 180$^{\circ}$ out of phase &
Electric dipolar and DEP, magnetic torque for circular orbits.\\
\\

\hline
\end{tabular*}

\end{table*}

\section{Results and Discussion}
Experiments on field-induced propulsion and assembly of colloids were performed in a homemade setup integrated with an optical microscope. Electric field activation-based experiments were performed by constructing a capillary cell which consisted of the sample (superparamagnetic polystyrene spheres of size 1 $\mu$m) sandwiched between two Indium tin oxide (ITO) coated cover slips (electrodes) separated by a parafilm spacer of thickness 150 $\mu$m, as shown in \textbf{Fig. 1}a. The ITO electrodes were placed in a skewed manner in order to take out electrical contacts without shorting the system. AC electric field of desired frequency and peak-to-peak voltage (Vpp) was given using a function generator. DC magnetic fields were applied using a Helmholtz coil, either placed along the x-y plane (for in-plane magnetic field activation) or along the z axis (for out-of-plane magnetic field activation), as shown in the figure (Fig. 1a). The entire setup was integrated with the sample stage of an upright microscope, and both single-field and multi-field experiments were performed.
\subsection{Structure formation under single field: electric or magnetic}
Before exploring complex assemblies under combined fields, it is essential to briefly discuss the roles of purely electric and purely magnetic fields in structure formation. The assembly of 1 $\mu$m polystyrene particles under the influence of an out-of-plane AC electric field depends on several factors, including amplitude and frequency of the applied field, zeta potential of the particle, and ion concentration in the solvent.\cite{barros2025assembly} At high applied field frequencies ($\sim$100 kHz-1 MHz), and a perpendicularly applied AC electric field (as shown in Fig. 1a), dipole-dipole interactions dominate, leading to the formation of vertical stacks or chains aligned with the field. However, at lower frequencies (5-20 kHz), electrohydrodynamic (EHD) flows become more significant than dielectrophoresis and interparticle dipole-dipole interactions, causing the particles to form quasi-planar crystalline or glassy structures\cite{barros2025assembly} near the substrate (electrode), as shown in Fig. 1b. In contrast, when a purely magnetic (DC) field is applied in the x-y plane, particles align into linear chains parallel to the direction of the field (Fig. 1c). The magnetic interaction in this case is purely dipolar in nature. The strength of the magnetic field dictates the number of particles in a given chain segment and the final chain length. 
\begin{figure}[!h]
\centering
  \includegraphics[width=6.3cm]{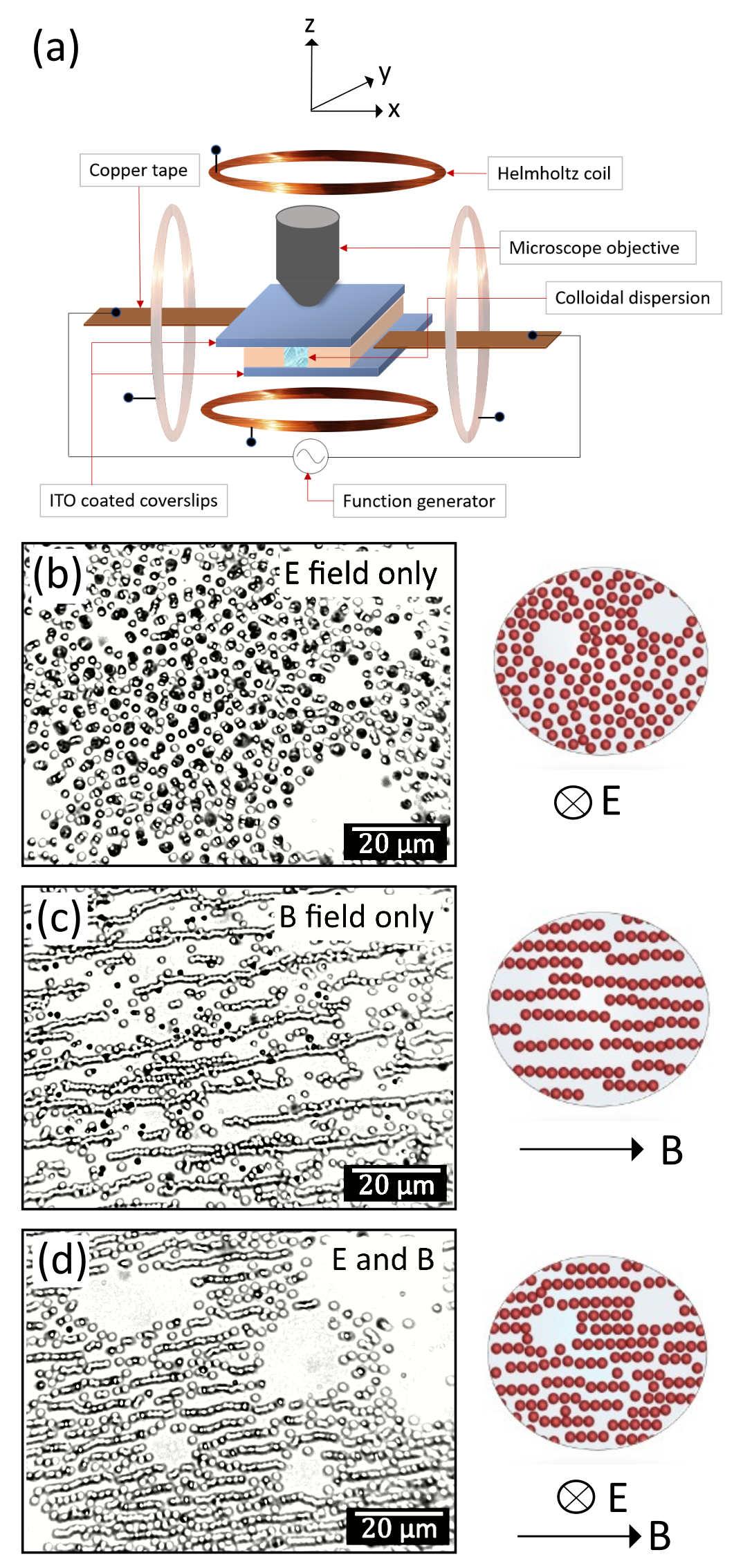}
  \caption{Structures produced from purely electric (E) and magnetic field (B) and one specific E field-B field combination: (a) The experimental setup showing the sample chamber, Cu contacts for applying the E field and Helmholtz coil configurations for applying in-plane and out-of-plane magnetic fields. (b) Under an out-of-plane electric field (AC) of frequency 10 kHz and peak-to-peak voltage (Vpp) 12 V, formation of quasi-planar, non close-packed clusters is observed. (c) A DC in-plane magnetic field of 0.62 mT produces linear chains due to the magnetic dipolar interactions between superparamagnetic polystyrene particles in the direction of the applied field. (d) A combination of an out-of-plane AC electric field and an in-plane magnetic field produces a two-level structure with large quasi-planar non close-packed clustered regions (`domains'), individually composed of linear chains. The schematic diagrams on the right of each picture bring out the structural variation across the three cases.}
  \label{fig1}
\end{figure}

A combined application of the electric (E) and magnetic field (B) will therefore produce complex structures where the configurational order created by both types of fields will either be reinforced/opposed by each other, or a completely new hierarchical structural rearrangement will be produced where the assembly at different length scales will be influenced by either of the two different fields. For example, in the previous case, if an out-of-plane AC electric field and an in-plane DC magnetic field are applied simultaneously, a two-level configuration will be produced (Fig. 1d) where chain segments composed of micrometre-sized polystyrene particles exist within large-sized ($\sim$ 50 $\mu$m) quasi-planar non-close packed clustered regions which we refer to as `domains' for convenience. Emergence of hierarchical structures in colloidal assembly has been computationally shown previously, where trimers formed from charge-stabilised magnetic colloids in the first stage act as building blocks for the formation of a hollow spherical structure in the final stage. \cite{DC_doublehierarchy} Similarly, cubic diamond and body-centred cubic crystals were predicted to be formed from triblock patchy colloids via a hierarchical assembly process.\cite{morphew_dwaipayan2018programming} In our case, while the electric field leads to non-close packed planar organisation of the particles into clustered domains in the mesoscale, the magnetic field imposes a microscale linear ordering (`chaining') inside those domains, thereby creating a two-level structure formed in two stages of reorganisation. Therefore, by subtly tuning the different parameters of the two types of fields, we can create a whole set of new structural configurations that would not be achievable by a single type of field alone. We demonstrate this in the following sections by first describing the structure formation under parallel electric (E) and magnetic (B) fields, where either a cooperation or a competition is observed depending upon the frequency regime of the applied E field. Next, we move on to assembly under crossed E and B fields, where more complex hierarchies start to develop.

\subsection{Electric and Magnetic fields in parallel: cooperation and competition}

\begin{figure*}[!h]
\centering
  \includegraphics[width=18.0 cm]{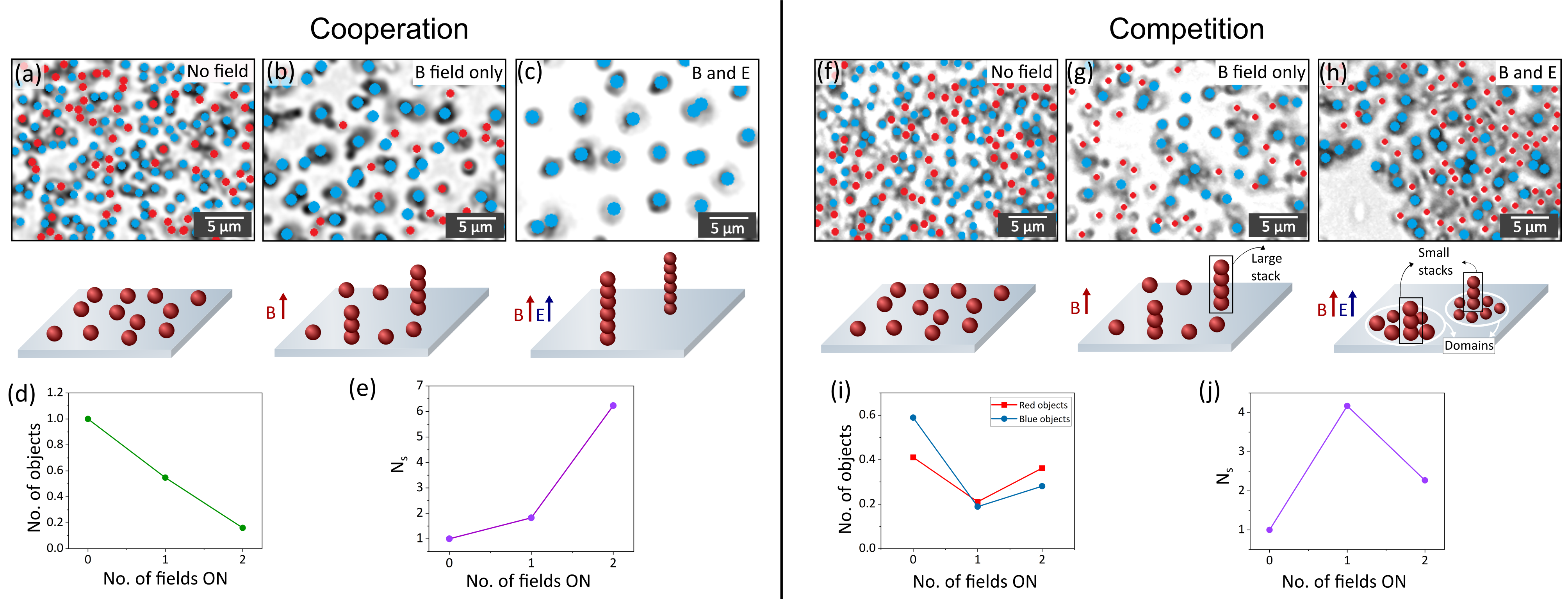}
  \caption{Cooperation and competition in structure formation under parallel electric and magnetic fields. (a)-(c) In the electric field frequency regime of 100 kHz-1 MHz, E and B fields act in cooperation. (a) Randomly dispersed particles in the bulk in the absence of any applied field. (b) Partial stacking of particles induced by an out-of-plane magnetic field of 0.62 mT. (c) The subsequent application of a 1 MHz AC electric field at peak-to-peak voltage (Vpp) of 20 V enhances the stacking process, incorporating all loose particles into longer columnar stacks. The red and the blue colours represent the particles in focus and out of focus on the viewing plane, respectively. The schematic diagrams representing the three cases are shown below the figures. (d) A plot showing the number of tracked objects (both red and blue, over an area of 6912 $\mu$m$^2$) plotted as a function of the number of fields acting on the particles. The total number of objects observed in each case is normalized with the maximum number of objects as seen in Fig. 2a (no field condition). The decreasing trend indicates a reinforced stacking of the particles under combined E and B fields. (e) Number of particles in each stack ($N_s$) plotted as a function of the number of fields in cooperation, showing a gradual increase from (a) to (c). In the lower frequency regime (5-20 kHz), the E and B fields compete with each other in structure formation as shown in (f)-(h). (f) Randomly dispersed particles in the absence of any applied field. (g) Partial stacking of particles induced by an out-of-plane magnetic field of 0.62 mT. (h) The subsequent application of a 5 kHz AC electric field at Vpp = 20 V breaks the stacks to form smaller stacks (shown in blue) and singlet particles (shown in red), both of which are part of a large local cluster. The schematics are given below the figures. (i) indicates the number of red and blue objects (normalized with respect to the maximum number of red and blue objects as seen in the zero field condition that is Fig. 2f), both of which show a sharp decrease followed by an increase from (f) to (g). (j) represents the number of particles per stack ($N_s$) which shows an increase followed by a decrease under an increasing number of fields, indicating that stacking produced by the B field is opposed by the E field, leading to the incomplete breakdown of larger stacks.}
\label{fig2}
\end{figure*}

Under the influence of parallel electric and magnetic fields, the resulting structure emerges from either cooperative or competitive interactions between the two fields. For a high frequency (100 kHz-1 MHz) out-of-plane E field and an out-of-plane B field in the range of 0.2---1.2 mT, large vertical columnar stacks of particles are formed via a cooperation between the electric and magnetic forces. In this frequency regime, the dipolar contribution of the electric force dominates, leading to vertical stacking of the particles in the direction of the field. If we represent the applied AC electric field with frequency $\omega$ as the real part of the complex phasor $\textbf{E}(t) = E_0 e^{i\omega t}\hat{e_z}$, then the time-averaged (electric) dipole-dipole interaction force on a particle at position \textbf{r} exerted by another particle residing at the coordinate origin is given as:\cite{barros2025assembly,Jingjing_Wu_openlattice}
 \begin{equation}
     \textbf{F}_{E}(\textbf{r})=\frac{3}{4}\pi\epsilon _m a^2\lvert{C_0}\rvert^2E_{rms}^2 \left(\frac{2a}{r}\right)^4 \left[(1-3\cos^2{\theta})\hat{r}+\sin{2\theta}\hat{\theta}\right]  
\label{fdip}    
\end{equation}

Here, $CM$ denotes the frequency-dependent complex Clausius–Mossotti (CM) factor, expressed as $CM=\frac{\epsilon_p^*-\epsilon_m^*}{\epsilon_p^*+2\epsilon_m^*}$ where $\epsilon_p^*$ and $\epsilon_m^*$ are the complex permittivities of the particle and the suspending medium, respectively and $C_0$ is the magnitude of the CM factor (see supplementary section 3). In this expression, $\epsilon_m$ is the medium permittivity, $a$ is the particle radius, $\textbf{r}$ represents the center-to-center separation between particles, $E_{rms}$ is the root-mean-square (rms) value of the applied electric field, and $\theta$ is the angle between the applied field direction and the interparticle axis. For $\theta = 0^\circ$, dipolar interactions lead to attraction between the particles, whereas for $\theta = 90^\circ$, the interaction is repulsive. It is clear from equation (1) that the electric dipolar force depends on the voltage ($\sim E_{rms}^2$) and the size of the particles ($\sim a^6$), due to which the force will be dominant for higher voltages and larger-sized particles. The frequency dependence comes from the Clausius-Mossotti factor (see supplementary section 3).
This electric dipolar contribution is reinforced by the application of the magnetic field, which also leads to out-of-plane chain formation due to the magnetic dipolar interactions. The magnetic dipolar force between two magnetized particles in a DC magnetic field $H$ is given as :\cite{griffiths2013introduction}
\begin{equation}
    \textbf{F}_{B}(\textbf{r})=\frac{3\mu_0m^2}{4\pi r^4}\left[(1-3\cos^2{\theta})\hat{r}+\sin{2\theta}\hat{\theta}\right] 
\end{equation}
where the magnetic dipole moment is given as $m=\frac{4}{3} \pi a^3 \chi H$. Here $\chi$ is the dimensionless volume susceptibility, $\mu_0$ is the magnetic permeability of free space and $a$ is the particle radius. We again assume that one particle is at the coordinate origin with the other particle being at a position $\textbf{r}$. The angle between the applied field direction and the inter-particle axis is $\theta$ as before. Similar to the electric dipolar force, this force is also dependent on the particle size ($\sim a^6$) and magnetic field strength ($\sim B^2$,) and therefore increasing these parameters would lead to more pronounced chains or stacks in the direction of the applied field.
In this multi-field assembly, the field that is applied second cooperates with the field applied first by adding any loose particles to the existing stacks and combining smaller stacks- thereby making very large stacks of particles as shown in \textbf{Fig. 2}a-c. Using ImageJ, we found the number of objects in the field of view for the three cases: case a: no-field (Fig. 2a), case b: B field only (Fig. 2b) and case c: B and E in parallel (Fig. 2c). The objects in focus in the viewing plane appear bright and are marked with a red dot in Fig. 2a-c, while the objects out of the plane appear dark and are marked with a blue dot. The total number of objects counted includes both the red and the blue dots. The addition of particles to an already existing stack is seen in Supplementary video V1 and is presented schematically below Fig. 2a-c. Note that it is easier to observe the structure formation from the video (V1) than from the still images (Fig. 2a-c). This also applies to the different structural configurations discussed in successive sections of the paper. The total number of objects (normalised by the maximum number of objects in the field of view) shows a sharp decrease from case a to case c (Fig. 2d), indicating reinforced stacking under combined out-of-plane E and B fields. The approximate number of particles in each stack was calculated by dividing the total number of objects in Fig. 2a by the total number of objects in Fig. 2c, assuming that all the particles in the viewing plane contributed to the stacking. An increase in the number of particles in each stack ($N_s$) (Fig. 2e) under an increasing number of fields also indicates a cooperation between the E and B fields.

\begin{figure}[!h]
\centering
  \includegraphics[width=8.3 cm]{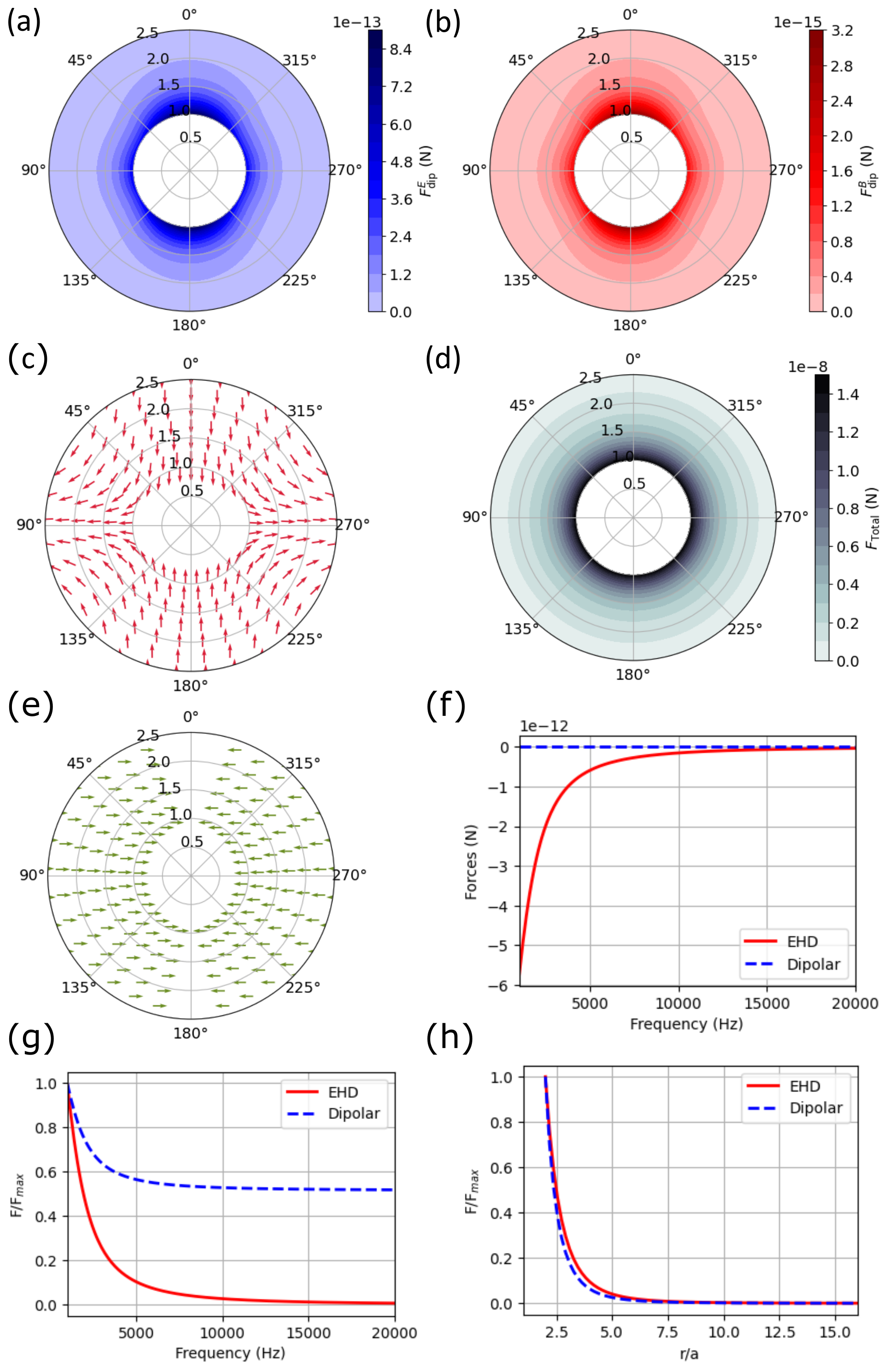}
  \caption{Quantification of interaction forces: (a) Electric dipolar force field generated by a source particle at 5 kHz, shown as a colour map of the force experienced by a test particle. The colour map is plotted on the x-z plane where 0$^{\circ}$ represents the direction of the z axis. (b) Magnetic dipolar force field, showing a similar behaviour, but the force value is two orders of magnitude smaller than the electric dipolar forces. (c) Corresponding vector plot of the force field illustrating the directionality of the dipolar interactions, showing attractive and repulsive zones parallel and perpendicular to the field directions, respectively. (d) Resultant force colourmap obtained from the superposition of electrohydrodynamic (EHD), electric dipolar and magnetic dipolar forces at a frequency 5 kHz. (e) The resultant vector diagram of the total force showing the dominance of EHD over dipolar forces at this frequency and a contractile EHD flow. The arrows are not shown near 0$^{\circ}$ and 180$^{\circ}$ as our equation for the tangential EHD force (equation (3)) cannot accurately estimate the force field in this region.  (f) A plot of EHD (red) and electric dipolar (blue dashed) forces as a function of frequency at a distance of 10 $\mu$m showing a drop in the EHD force magnitude as the frequency increases. (g) A plot of the scaled force $F/F_{max}$ as a function of frequency clearly brings out the fall in EHD at higher frequencies, and (h) The scaled ($F/F_{max}$) EHD and dipolar forces fall off similarly with distance at a frequency of 5 kHz.}
\label{fig3_new}
\end{figure}

\begin{figure*}\centering
  \includegraphics[width=18.0cm]{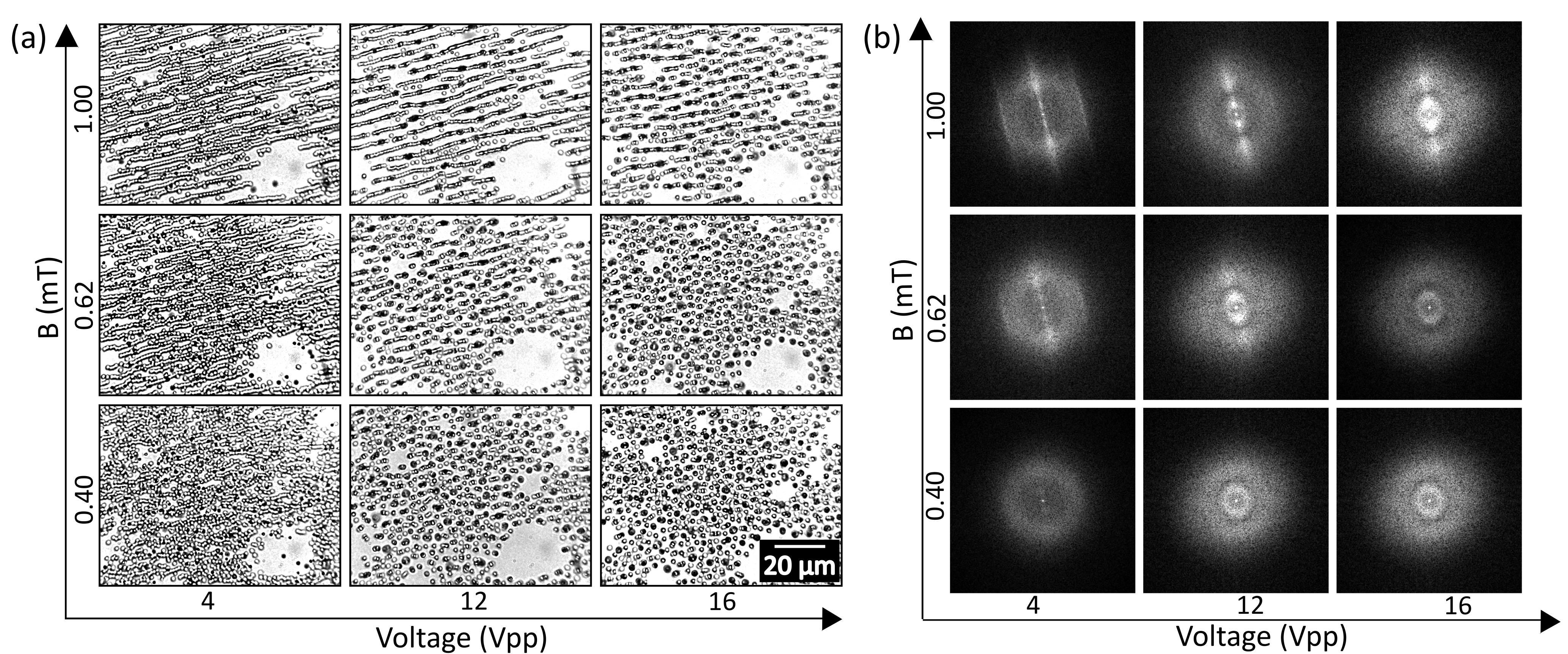}
  \caption{Structural configurations formed under crossed E and B fields as a function of the magnetic field strength and peak-to-peak voltage of the electric field. (a) Optical micrographs at a constant E field frequency of 10 kHz show predominance of chain formation at higher magnetic fields and formation of non close-packed colloidal clusters at higher electric field values. In the intermediate regime, the electric and magnetic forces reach a subtle balance, leading to a hierarchical arrangement where short chain segments constitute clustered `domains' separated by large gaps. (b) Fast Fourier Transform (FFT) of the images bring out the appearance and disappearance of oriented configurations and structural order for particular values of E and B.}
\label{fig3}
\end{figure*}
At lower electric field frequencies (5-20 kHz), where particles typically adopt planar configurations,\cite{barros2025assembly} the competing influences of the E and B fields lead to the formation of stable yet complex structures. The competition here arises because in this frequency regime, the electrohydrodynamic force\cite{ristenpart2004assemblypre,ristenpart2007electrohydrodynamic} dominates over electric dipole-dipole interactions, leading to structure formation on the horizontal (x-y plane) while the magnetic forces are still dipolar only, leading to stacking along the z axis. The tangential EHD force field surrounding a spherical colloidal source particle situated close to an electrode is given
as:\cite{ma2015inducing, ristenpart2004assemblypre}
%\begin{equation}
 %   \vec{F}_{ij}^{EHD} = - \frac{A_j a^4 h_i \vec{r}_{ij}}{r_{ij}^5 h_j^2}
%\end{equation}

\begin{equation}
F_{EHD} = \beta \left(\frac{9}{2}\pi \varepsilon_m a\right)\,\kappa h E_{\mathrm{rms}}^2
\frac{
C_0' + \bar{\omega} C_0''
}{
1 + \bar{\omega}^2
}
\frac{\left(\frac{r}{a}\right)}{\left[1 + \left(\frac{r}{a}\right)^2\right]^{5/2}}
\end{equation}

where $\bar\omega = \frac{\omega h}{2 \kappa D}$ is the scaled frequency, $\kappa^{-1}$ is the Debye screening length, D is the ion diffusivity, $h$ is the separation between the two electrodes and $\beta=0.04$ is a scaling factor.\cite{ma2015inducing} As discussed in our earlier work,\cite{barros2025assembly} the tangential flow velocity and hence the EHD force is dependent on the frequency ($\sim \omega ^{-1}$) of the applied E field. At lower frequencies, therefore (5-20 kHz), the EHD force is dominant over the dipolar force, thereby forming the quasi-planar non close-packed structures.

In this regime, the system reaches a balance, conforming to both fields while partially preserving the structural features created by each. For instance, while the low-frequency electric field (5–20 kHz) promotes quasi-planar non close-packed clustering, the out of plane magnetic field leads to stacking along the z-axis, leading to a competition between the two forces. The particles in this case organise into smaller stacks embedded within the larger planar domains created by the electric field. This is seen from Fig. 2f-h and supplementary video V2. Here also the particles in focus on the viewing plane are marked with a red dot and those out of focus are marked with a blue dot. In zero field condition, a large number of blue and red dots are seen (Fig. 2f), whose numbers get reduced on application of the B field due to vertical stacking (Fig. 2g) and rise again (Fig. 2h) due to breaking of the vertical stacks into a) smaller stacks (marked in blue) and b) individual spread out particles on the x-y plane (marked in red). The schematic diagrams given below Fig. 2f-h, illustrate each of these stages. The reduction in the number of both blue and red objects on application of the B field, and subsequent increase on application of both E and B fields, is clearly seen from Fig. 2i. The number of particles in each stack ($N_s$) in Fig. 2h was calculated by first subtracting the number of red objects (particles in plane) in Fig. 2h from the total number of objects (Fig. 2f), and then dividing this by the total number of blue objects (number of stacks) in Fig. 2h. $N_s$ was observed to increase and subsequently decrease (Fig. 2j), thereby bringing out the competition between the E and B fields. This is a hybrid arrangement where, even in a competition, neither field completely dominates, but rather, a compromise between the two is established. 

We performed a quantitative analysis of the electric and magnetic dipolar and EHD forces in the two dimensional x-z plane to examine the aspects of cooperation and competition. The details of our calculations are given in supplementary section 3. Using equations (1) and (2), the electric and magnetic dipolar force fields generated around a 1 $\mu$m source particle were estimated, as shown in Fig. 3a and b, respectively. In these figures, $0^{\circ}$ represents the z direction along which the (AC) electric field or the (DC) magnetic field is applied, while the horizontal direction represents the x axis. For the electric dipolar case, at a frequency of 5 kHz, the maximum force field ($\sim10^{-13}$ N) is observed at the edges of the particle along the z direction, which falls off as $1/r^4$, where $r$ is the distance from the center of the source particle. The magnetic forces were estimated by considering 26.5\% of magnetite (as per the particle manufacturer's datasheet) embedded within the polystyrene matrix. The magnetic dipolar forces were estimated to be two orders of magnitude smaller ($\sim10^{-15}$ N) than the electric dipolar forces at a maximum magnetic field of 1 mT as used in our experiments. The fact that even such small magnetic forces lead to substantial chaining along the field direction and significant response to the applied field\cite{haque2023propulsion, haque2025high, demirors2017colloidal, zhu2021synthesis, zhu2025reconfigurable} arises from the fact that the local magnetic field gradient experienced by a particle in the vicinity of multiple magnetized particles may be much higher than the applied field. The vector (quiver) plot of the combined electric and magnetic dipolar forces is given in Fig. 3c, which as expected, shows attraction along the vertical z direction and repulsion along the horizontal x direction. For the calculation of the EHD forces, equation (3) was used, which showed a force of the order of magnitude of $\sim10^{-8}$ N at an electric field frequency of 5 kHz. A combined plot (Fig. 3d) of the three forces (electric and magnetic dipolar, and EHD) at 5 kHz therefore shows a complete dominance of the EHD forces over the dipolar forces. The vector plot in Fig. 3e indicates that the EHD flow is contractile, that is directed towards the particle, with an overall negative sign. The negative sign arises from the fact that for a contractile flow, the term $\frac{
C_0' + \bar{\omega} C_0''
}{
1 + \bar{\omega}^2
}$ is negative. A comparative plot of the electric dipolar and EHD forces at a distance of 10 $\mu$m and an angle of $90^\circ$ from the source particle is given in Fig. 3f. Firstly, we see that while the dipolar force is positive, the EHD force is negative. Secondly, the EHD force dominates in the low frequency regime (1-5 kHz), but shows a sharp fall at higher frequencies and becomes comparable or smaller than the dipolar force beyond 20 kHz. This explains the `competition' effect seen at lower frequencies in structure formation ( Fig. 2f-j), while at higher frequencies, the electric and magnetic dipolar forces dominate, leading to cooperation between them (Fig. 2a-e). The dominance of the dipolar forces over EHD forces at higher frequencies is more succinctly brought out by plotting the ratio $F/F_{max}$ ($F_{max}$ being the maximum value of the specific type of force) in Fig. 3g. It is clearly seen that the dipolar force tends to dominate over EHD beyond 10 kHz which leads to the formation of cooperative chains along the z axis at higher frequency ranges. Lastly, a plot of $F/F_{max}$ of the dipolar and EHD forces vs the scaled distance ($r/a$ where $a$ is the radius of the particle) from the source particle at a frequency of 5 kHz, shows that the dipolar and EHD forces both fall off with distance in a nearly similar manner. It should be noted that our calculations are based on analytic expressions of dipolar and EHD forces between a pair of colloidal particles,\cite{ma2015inducing, ristenpart2004assemblypre, griffiths2013introduction, morgangreen2003ac} and an accurate estimate of the forces can only be made by numerical simulations involving the behaviour of the fields near the polarizable electrode surfaces and the contributions from the involvement of multiple particles. We believe that this involves an extensive, in-depth examination and plan to follow it up in future work. 

\subsection{Crossed electric and magnetic fields: multi-scale organisation}
In contrast to the cooperative or competitive effects of the electric and magnetic forces as observed in the case of the parallel E and B fields, a whole new picture emerges when the two fields are applied in perpendicular directions. A combination of an in-plane magnetic field and an out-of-plane AC electric field produces a two-level structure, where the contributions of both fields are observed at different length scales. For example, initially, when a low-frequency (10 kHz) out-of-plane electric field is applied, the particles assemble into a quasi-planar, non close-packed structural configuration consisting of clustered regions or `domains' which are tens of micrometers in size, and are separated by large empty spaces (Fig. 1b). Following this, the application of an in-plane magnetic field of 0.62 mT rearranges the individual particles within these planar domains into chains without disrupting the overall planar order (Fig. 1d).

\begin{figure*}[h]
\centering
  \includegraphics[width=18.0 cm]{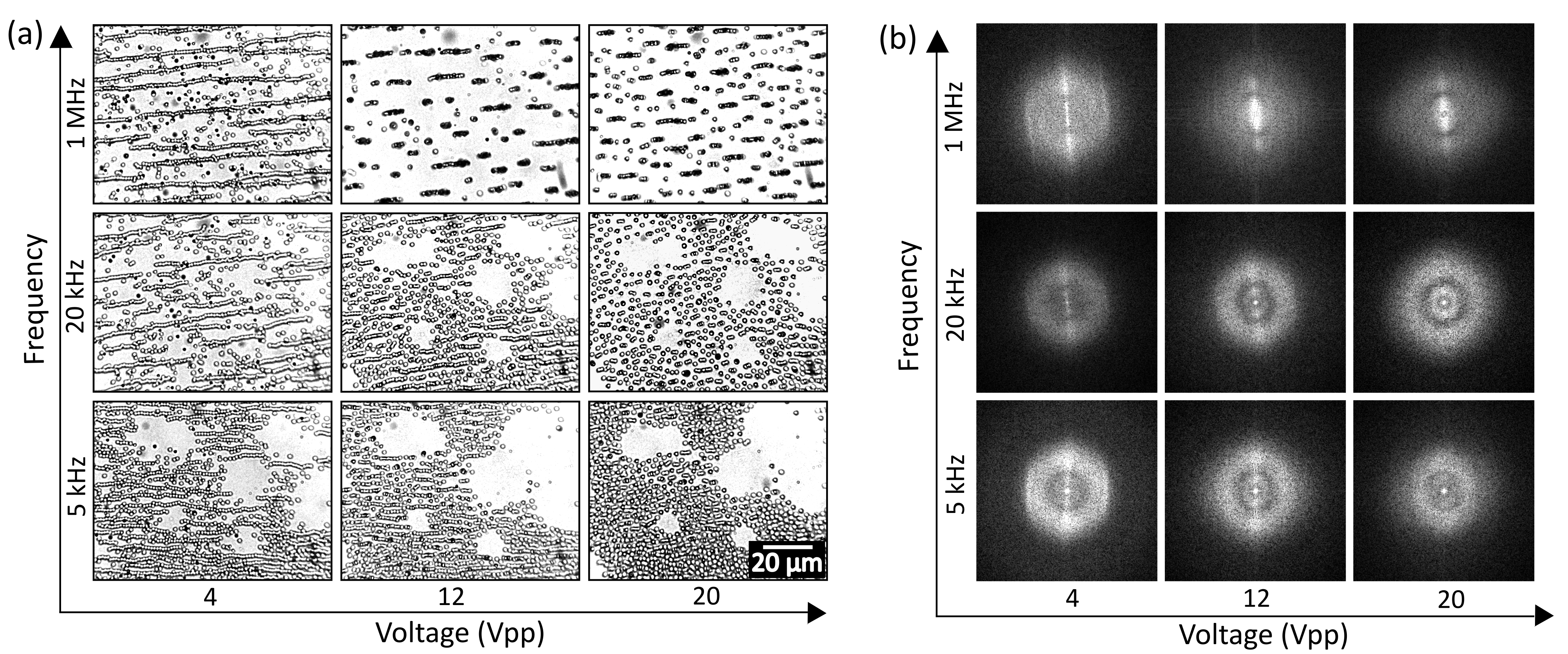}
  \caption{Structural configurations formed under crossed E and B fields as a function of the frequency and peak-to-peak voltage (Vpp) of the electric field. (a) Optical micrographs at a constant magnetic field strength of 0.62 mT show predominance of chain formation at higher frequencies and formation of non close-packed colloidal clusters at high Vpp values. In the intermediate regime, the attractive and repulsive forces originating from dipolar and electrohydrodynamic forces reach a subtle balance. This leads to a hierarchical arrangement where short, separated chain segments constitute clustered `domains' separated by large gaps. (b) Fast Fourier Transform (FFT) of the images indicate the appearance and disappearance of oriented configurations and structural order for particular values of E and B.}
\label{fig4 }
\end{figure*}

The structural configurations of the hierarchical assemblies depend on a subtle interplay between the electric and magnetic forces. The electric forces, which are dipolar, dielectrophoretic and electrohydrodynamic, depend on the applied field frequency and amplitude, while the strength of the applied magnetic field is varied by changing the voltage applied to the Helmholtz coils. Fig. 4 and supplementary Fig. S1 present the range of structural configurations produced under an out-of-plane electric field (10 kHz) and an in-plane magnetic field, as the magnetic field strength and the electric field’s peak-to-peak voltage (Vpp) are varied. For low values of the E field voltage (4 Vpp), an increase in the B field increases the length of the chains (Fig. 4a). However, as the electric field voltage is increased beyond a certain threshold, magnetic dipolar interactions are overpowered by the electrohydrodynamic (EHD) forces, leading to the formation of shorter chain segments. For the maximum values of E field voltage (16 Vpp) and B field (1 mT) in our experiment, we observe the formation of short chain-like segments constituting larger area `domains'. The morphology of these structures is therefore fully controllable by tuning the electric and magnetic field parameters.  If one of the forces dominates over the other, the structure moves either towards long chains spanning several microns (B field dominating) or clusters composed of singlet particles arranged in non close-packed configurations (E field dominating). To quantify the structural configurations, we took the Fast Fourier Transform (FFT) \cite{lotito2020pattern} of each image as shown in Fig. 4b and supplementary Fig. S2. At high B field and low E field values (1 mT and 4 Vpp), the FFT shows a clear anisotropy with bright lines oriented in a direction perpendicular to the long chains in Fig. 4a. While at low B field and high E field values (0.4 mT and 16 Vpp),  a diffused ring pattern is observed, indicating the destruction of linear order and formation of local clustered zones. In all other intermediate cases, we have features of both ordered (chaining) and disordered (clustered) configurations, with the hierarchy being most evident in the strongest E field-B field case (16 Vpp and 1 mT).

To explore frequency-tunable structural transitions, we studied particle assemblies under a constant in-plane magnetic field of 0.62 mT, while varying the frequency and voltage (Vpp) of an out-of-plane AC electric field (Fig. 5 and supplementary Fig. S3). At high frequencies ($\sim$ 100 kHz–1 MHz), and at low voltages (4 Vpp), the magnetic field dominates, resulting in the formation of long chains (Fig. 5a). The corresponding FFT (Fig. 5b and supplementary Fig. S4) exhibits bright lines oriented perpendicular to the chains, confirming their linear order. In this frequency regime, electric dipole–dipole interactions are stronger than EHD effects, favoring stacked arrangements, but the stacking is overpowered by the magnetic interactions as the voltage is low. Increasing the electric field voltage enhances this stacking tendency, leading to fragmentation of the long chains into smaller chains and chain segments with stacked particles. At lower frequencies ($\sim$ 5–20 kHz), where EHD interactions become significant, the long magnetic chains destabilise and reorganise into quasi-planar, non close-packed clusters. Increasing the amplitude in this regime further amplifies EHD effects, breaking chains into progressively smaller fragments. At the lowest frequency and highest voltage examined (5 kHz, 20 Vpp), the assemblies consist of the shortest chains embedded within clustered domains. This structural transition is reflected in the FFT, which shows a diffused ring pattern indicative of isotropic arrangements (Fig. 5b). Interestingly, the chain lengths, which are attributed to magnetic field strength, were controllable by tuning only the frequency and amplitude of the AC electric field while maintaining a constant magnetic field strength.

\begin{figure}[!h]
\centering
  \includegraphics[width=7.0 cm]{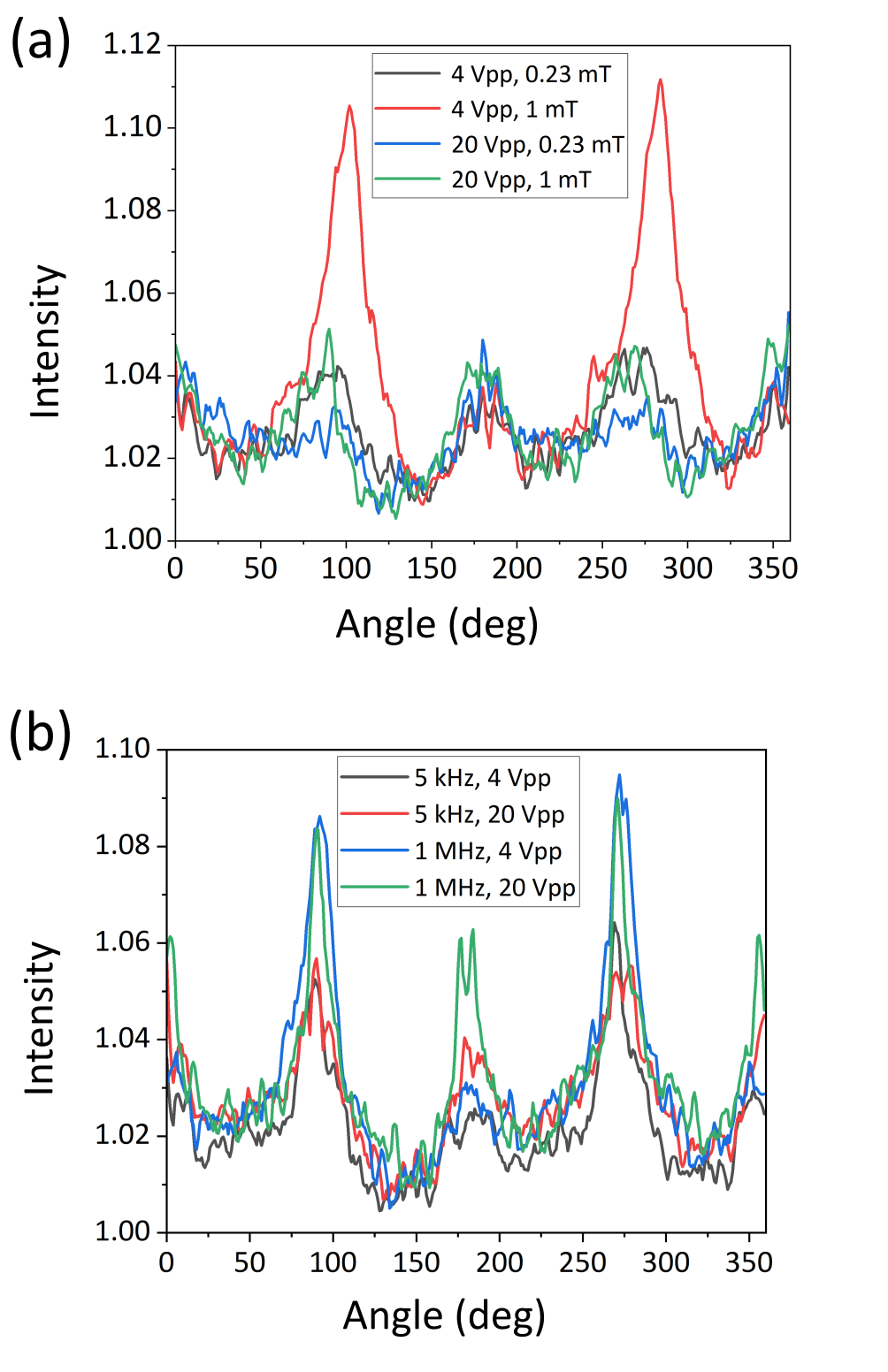}
  \caption{Anisotropy in structural configurations quantified using Fast Fourier Transforms of the images. Angular intensity profiles were obtained from FFT images, where the intensity is summed along all points on a radius at a given angle for (a) the constant-frequency case (corresponding to Fig. 4) and (b) for the constant magnetic-field case (corresponding to Fig. 5). The anisotropy in both cases are clearly observed with case (b) showing overall a larger anisotropy than case (a). Tuning of the field parameters leads to a reduction of the anisotropy. }
\label{fig5}
\end{figure}

\begin{figure*}[!h]
\centering
  \includegraphics[width=18.0 cm]{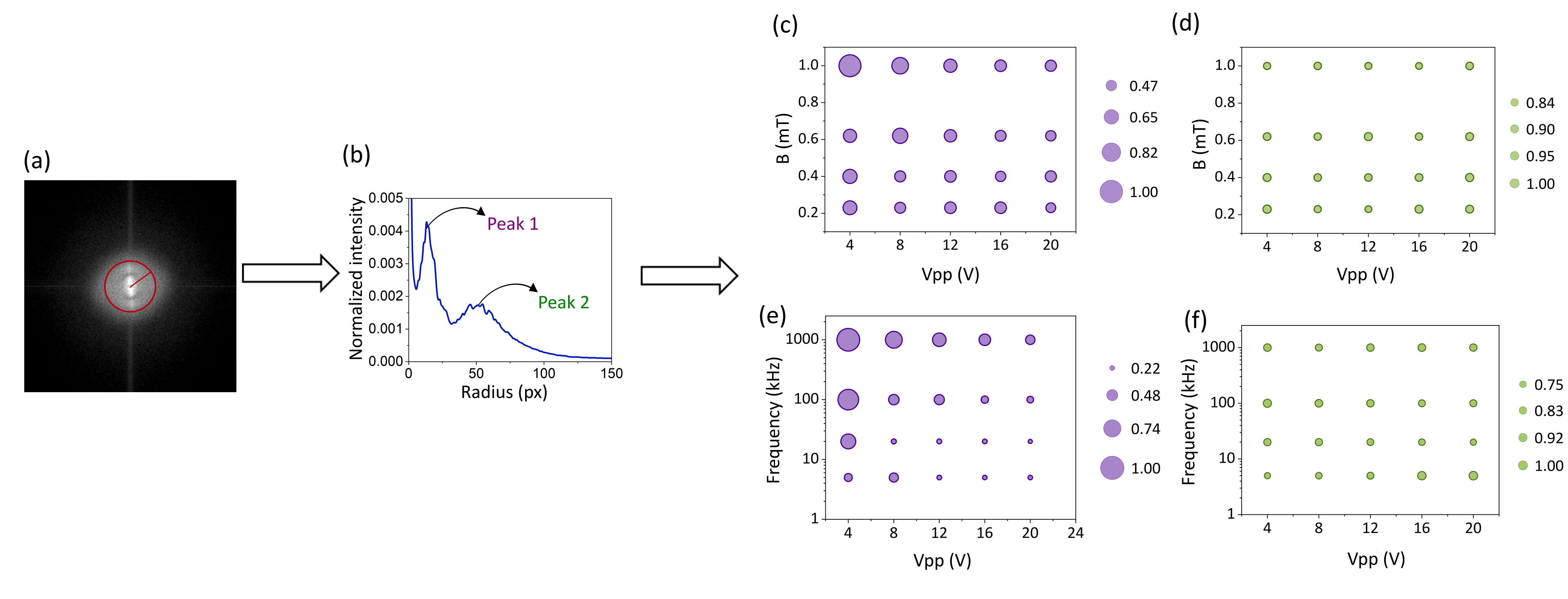}
  \caption{Analysis of relevant length scales from FFT images. (a) A representative FFT image showing the calculation of the radial intensity profile by summing the intensity values over a circumference at a given radius. (b) From the profile plot, the peak positions of the two major peaks are determined and converted to length scales in micrometers. We denote the peaks as peak 1 (marked in magenta) and peak 2 (marked in green) as shown in the figure. While peak 1 corresponds to a higher length scale, peak 2 corresponds to a lower length scale. Under a constant frequency of 10 kHz, (c) and (d) represent peak positions for peak 1 and peak 2, respectively, as a function of B field strength and Vpp of the E field. In these bubble plots, the size of the circles is proportional to the magnitude of the length scale corresponding to the peak position. (e) and (f) are similar bubble plots for the variation of frequency and Vpp of the E field while B field is kept constant at 0.62 mT. While (c) and (e) show strong variations in the relevant length scale (extracted from peak 1) across different values of the field parameters, the length scale calculated from peak 2 remains nearly unaffected as shown in (d) and (f).}
\label{fig6}
\end{figure*}

\begin{figure*}[!h]
\centering
  \includegraphics[width=18.3cm]{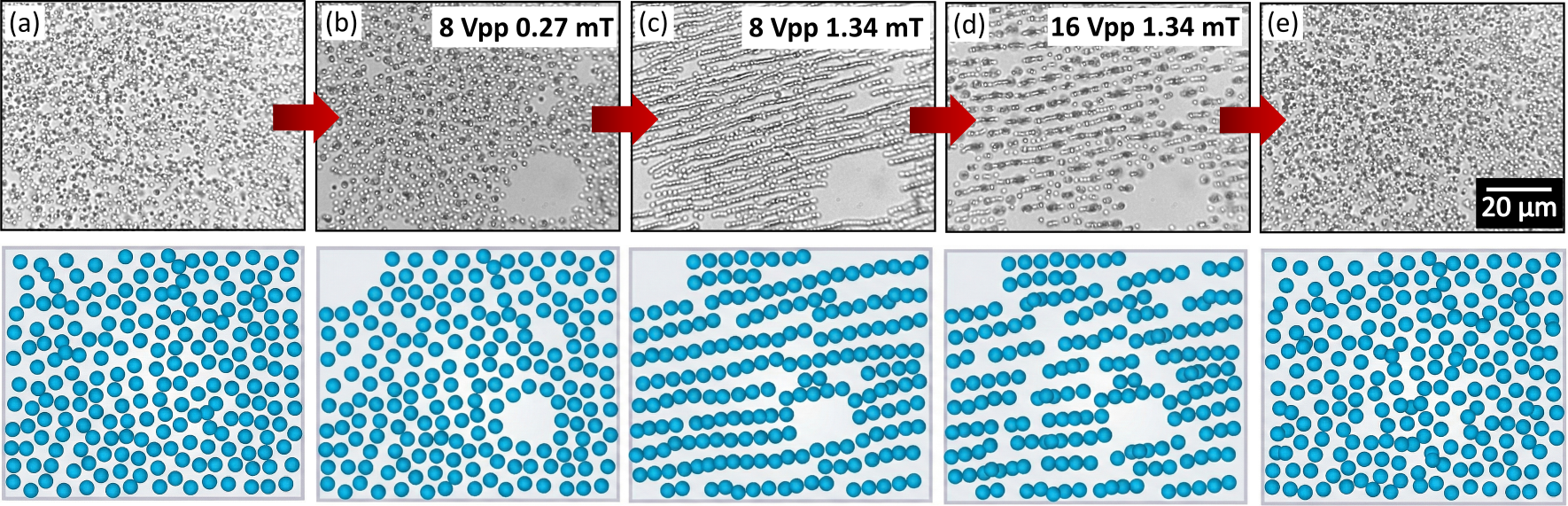}
  \caption{Explicit demonstration of reconfigurability under crossed electric and magnetic fields. By choosing the correct values of the frequency and amplitude of the electric field, the strength of the magnetic field, and the sequence of application, a whole series of structures can be reversibly obtained from the same sample. While under zero field the particles were dispersed in 3D (a), application of an electric field of 8 Vpp and a small magnetic field (0.27 mT) led to the formation of quasi-planar clustered 'domains' (b). Increasing the  magnetic field strength to 1.34 mT led to the emergence of long chains (c) without breaking the overall spatial arrangement, while increasing the voltage led to the disintegration of these chains into smaller segments and a few small clusters along the z axis. Switching off all the fields led to the return of the bulk dispersed configuration. Any of the above demonstrated structures can be reversibly obtained only by tuning of the E and B fields. The schematic diagrams are shown below the images to illustrate the structural configurations succinctly.}
\label{fig8}
\end{figure*}

For both the constant frequency (Fig. 4) and constant magnetic field (Fig. 5) cases, the FFTs of the images bring out the structural evolution and the appearance and disappearance of oriented structures under changing field parameters. To quantify the anisotropy observed in the FFT images in Fig. 4b and 5b, we plotted the angular intensity profiles by summing the intensity at all points on the radius at a given angle. Fig. 6a and b show the angular intensity plots for the constant frequency case (Fig. 4) and for the constant magnetic field case (Fig. 5), respectively. For clarity of the graphs, only four extreme situations are represented in each case. For the constant frequency case, two sharp peaks are observed at approximately 90$^\circ$ and 270$^\circ$ along with one small peak at 180$^\circ$ for the lowest voltage (4 Vpp) and the highest magnetic field (1 mT) values (Fig. 6a). This corresponds to the longest chains for the constant frequency case, as shown in Fig. 4. As the magnetic field decreases, or the voltage increases, this anisotropy gets reduced and the peak heights become comparable. Similarly, for the constant magnetic field case (Fig. 6b), maximum anisotropy is observed for the highest frequency, lowest voltage (1 MHz, 4 Vpp) case, and the anisotropy is reduced gradually with higher values of voltage and lower values of the frequency. This is also in resonance with our observations shown in Fig. 5a. The overall higher degree of anisotropy in the graphs in Fig. 6b as compared to Fig. 6a, indicates the higher predominance of linear chains for the constant magnetic field case as compared to the constant frequency case. In the constant frequency case, significant anisotropy is only observed in a small regime of magnetic field and voltage values.

Other than characterisation of the structural anisotropies in the configurations, FFTs can also give us information about the relevant length scales existing in the system. To explore this, we took radial intensity profiles from the FFTs as shown in Fig. 7a. The intensity at a given radial distance from the centre in the FFT image was found by summing the intensity over a circumference corresponding to that radius. This intensity, after normalization, was plotted as a function of the radial distance (Fig. 7b). In all radial intensity plots, two distinct peaks (peak 1 and peak 2) were observed at two different radius values (Fig. 7b). These radius values from the FFT images were then converted into real space corresponding length scales using equation (4) as described in the methods section. For both the constant frequency (Fig. 4) and constant magnetic field (Fig. 5) cases, the length scales obtained from peaks 1 and 2 were calculated for a wide range of values of the variable parameters (B-Vpp and frequency-Vpp, respectively). The results are represented as bubble plots where the bubble size is proportional to the extracted length scale for either the constant frequency (Fig. 7c-d) or the constant magnetic field case (Fig. 7e-f). We suggest that while peak 1 (Fig. 7c and Fig. 7e) indicates a length scale that is an amalgamation of the inter-chain spacing and chain segment length, peak 2 (Fig. 7d and f) represents a shorter length scale comparable with the particle size. While a larger chain spacing leads to closely spaced peaks along the vertical axis of the FFT image, a shorter chain length leads to a spreading of the peaks along the horizontal axis of the FFT. An averaging over a circular profile of a given radius value, therefore, leads to a peak (peak 1) in the radial intensity profile and a corresponding length scale that contains the information of both chain segment length and inter-chain spacing. Fig. 7c provides a direct visualization of how this characteristic length scale systematically decreases as the electric field strength (Vpp) is increased, as well as when the magnetic field strength is decreased for the constant frequency case. For the constant magnetic field case, Fig. 7e quantitatively illustrates that this characteristic length decreases with an increase in the electric field strength (Vpp) and increases with increasing frequency of the E field. 
This quantitative analysis not only confirms the qualitative observations from the micrographs (Fig. 4 and 5), but also enables a systematic comparison across different experimental conditions. The length scale calculated from peak 2 (Fig. 7d and f) is nearly unaffected across different values of the field parameters, and therefore, we assume it is related to the size of the particles rather than any field-assembled structural configuration.

\subsection{Sequence matters}
Our experiments clearly demonstrate the structural reconfigurability of field-activated colloidal assemblies by fine-tuning of the field parameters. The degree of reconfigurability is brought out explicitly in Fig. 8 and supplementary video V3, where, by tuning the E and B field parameters, the same sample is used to reversibly produce a range of structural configurations. However, we also observed that the sequence of application of the E and B fields plays a significant role in the final structural configuration. The field that is applied first plays a dominant role in determining the overall arrangement. For example, when the E field (10 kHz, 12 V) is applied first, quasi-planar non close-packed clusters form (Fig. 9a and supplementary video V4). On application of the B field after this, the individual particles inside these clusters reconfigure into short chain segments (Fig. 9b). When the magnetic field is applied first, long chains form spanning the field of view (Fig. 9c and supplementary video V5). On application of a subsequent electric field, the EHD interactions fragment the chains to produce domain-like areas with clustered zones separated by large gaps (Fig. 9d). Interestingly, the chain segments formed for case 2 (B first, E second) are on average longer than the chain segments formed for case 1 (E first, B second). We quantified the difference between the two cases by calculating a characteristic length scale in a similar manner as was done for Fig. 7c and e. These length scale values are plotted in Fig. 9e for both cases. The plot can be divided into two distinct regions: region 1, where only the first field is applied and region 2, where both the first and second fields are simultaneously applied. From the plot, we see a deviation in the characteristic length between case 1 (plotted in red) and case 2 (plotted in blue), with case 2 having approximately longer lengths under both E and B fields. The importance of the sequence of application of fields arises from the fact that the first field initially creates a structural order, which is broken and reconfigured by the second field, but this reconfigurability depends on the initial configuration. This means that it is energetically less expensive to break longer chains into shorter chain segments (as in case 2) than to reorganise individual particles inside a large domain into chain segments (case 1). This highlights the asymmetric influence of the two fields, where the initial conditions set by the first field shape the subsequent assembly dynamics. 

\begin{figure}[!h]
\centering
  \includegraphics[width=7.3cm]{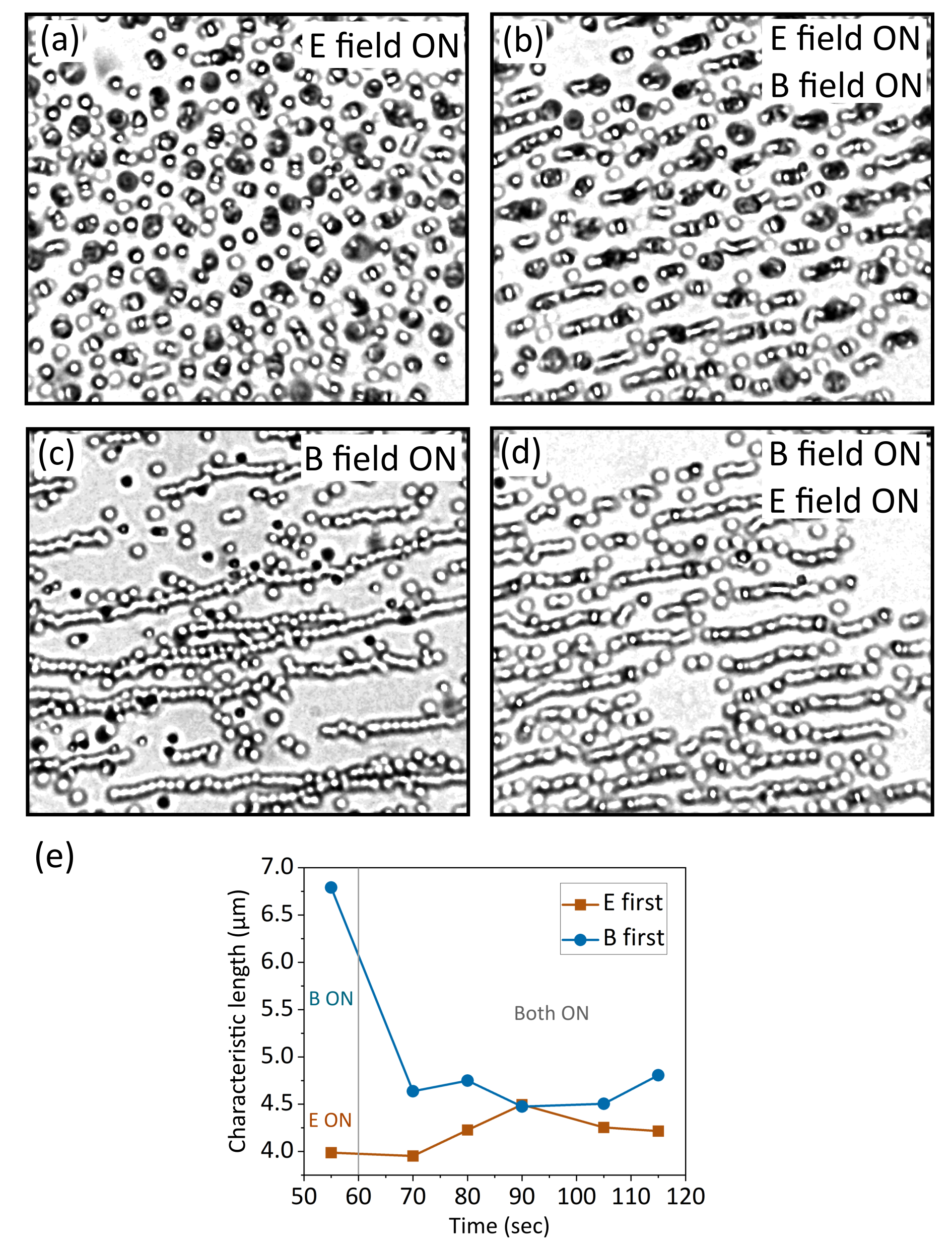}
  \caption{Sequence of application of fields directly influences structure formation: (a) When an out-of-plane E field (f = 10 kHz, Vpp = 12 V) is applied first, quasi-planar, non close-packed clusters form initially. (b) On subsequent application of an in-plane magnetic field of 0.62 mT, the individual particles in each domain rearrange into shorter chain segments and stacked chains by rearrangement of the particle positions. (c) An initial application of an in-plane magnetic field produces long chains along the field direction. (d) Subsequent application of an out-of-plane electric field leads to the fragmentation of chains into longer, but planar segments as compared to (b). (e) The characteristic length scale calculated from peak 1 in FFT for case 1: E applied first (a-b) and case 2: B applied first (c-d) is plotted as a function of time. The plot clearly shows the small but distinct difference in the structural configuration for the two cases, and therefore the importance of choosing the correct sequence of fields.}
\label{fig7}
\end{figure}

\subsection{Towards complexity: binary systems}
\begin{figure}[!h]
\centering
  \includegraphics[width=4.3cm]{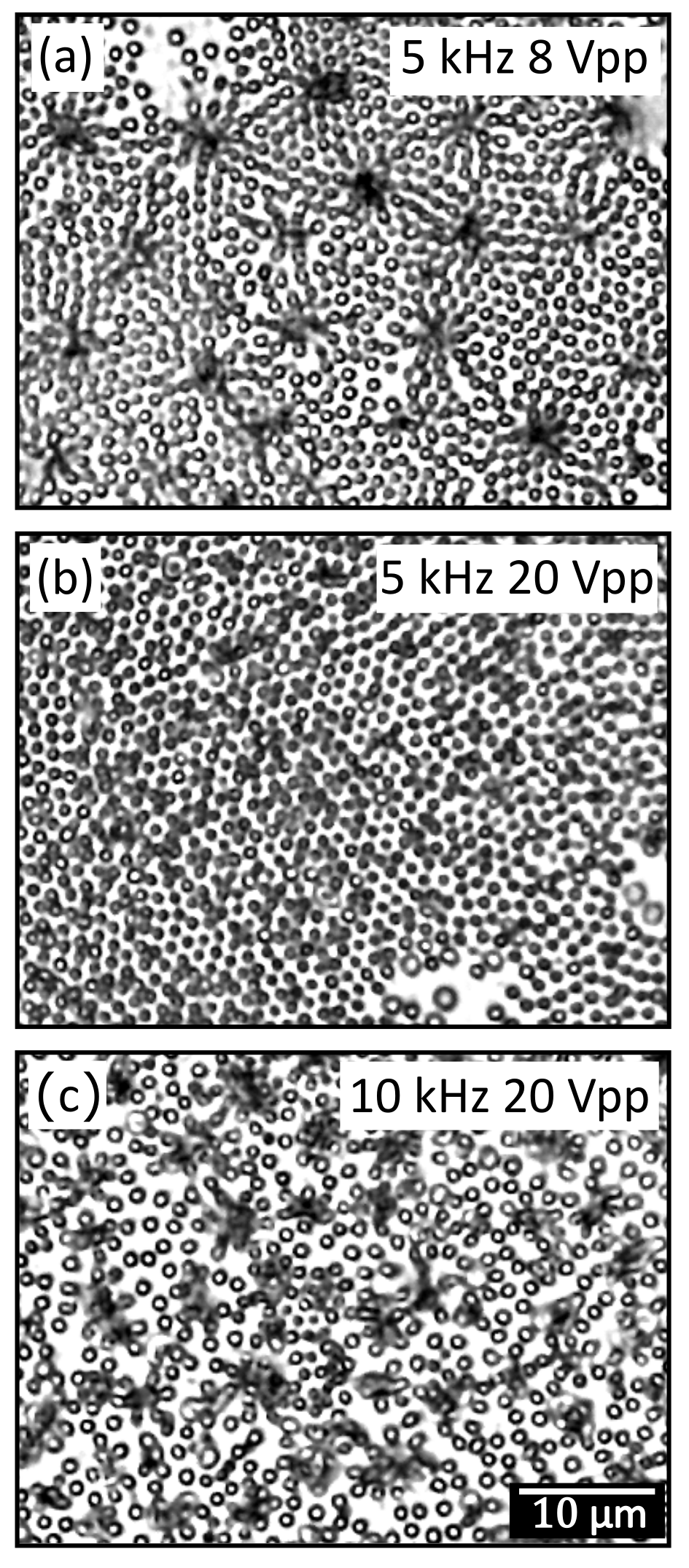}
  \caption{Formation of complex structures in binary particle systems under out-of-plane electric and magnetic fields. (a) Under an E field of f = 5 kHz, Vpp = 8 V, and B field of 1.16 mT, a binary mixture of magnetic and non-magnetic particles (both 1 $\mu$m in size) assembles into a particular configuration where the magnetic particles form `poles' and the non-magnetic particles form radial bridges between the poles. (b) An increase in the peak-to-peak voltage to 20 V brings more particles from the poles into the plane, creating many smaller stacks, with an ordered arrangement of particles bridging these stacks. (c) On increasing the frequency to 10 kHz at the same Vpp (20 V), the radial bridges are destroyed, leading to the formation of poles again with a random distribution of non-magnetic particles in between.}
\label{fig9}
\end{figure}
Beyond studying the assembly of spherical single-type spheres under multiple field excitations, we sought to explore the assembly behaviour of binary particle systems under the same combination of fields to introduce one more level of complexity. A binary mixture of 1 $\mu$m non-magnetic polystyrene particles and 1 $\mu$m superparamagnetic particles was prepared in a 1:1 ratio, maintaining similar number densities as before. When subjected to parallel electric and magnetic fields in the z direction (f = 5 kHz, Vpp = 8 V, B = 1.16 mT), the particles assembled into distinctive `pole'-like structures with radially arranged particles bridging the poles, as shown in Fig. 10a. The poles consisted of stacks of magnetic particles aligned along the magnetic field direction, as confirmed by increasing the field strength and observing the particles that responded to the field upon switching it off. The concentration of the magnetic particles at the poles is further emphasised by the fact that, in addition to the quasi-planar clusters created by the electric field, vertical stacks of particles also appear outside the clustered regions whenever a magnetic field is applied along the z-direction (see supplementary video V6). Because the experiment is performed at an electric-field frequency of 5 kHz, where EHD interactions dominate and favour the formation of planar or quasi-planar assemblies, the emergence of these vertical stacks cannot be attributed to the electric field. The simplest explanation is the action of the out-of-plane magnetic field, which induces vertical stacking of the magnetic particles of the binary mixture. On the other hand, the radial bridges connecting the poles were composed of non-magnetic particles along with some superparamagnetic particles. This suggests a competing interplay between the two external fields: while the magnetic interactions favour stacking along z, the electric field drives planar structuring. As a result, the system reaches a compromise, forming an intricate radiating or `flowery' pattern where some magnetic particles remain embedded within the planar domains rather than solely at the poles. Keeping the magnetic field constant at 1.16 mT, we tuned the geometry of the structures by varying the frequency and voltage of the E field. Under a constant frequency of 5 kHz, an increase in the voltage from 8 V to 20 V strengthens the EHD interactions, bringing more particles from the poles into the plane, thereby creating many smaller stacks (Fig. 10b). Also, the particles on the plane show an ordered, non close-packed arrangement in between the stacks. Increasing the frequency at the same voltage (20 Vpp) leads to a decrease in the EHD forces, therefore again reinforcing the stacking in z due to the existing B field. This leads to the formation of poles again, but the particles bridging the poles now reach a disordered arrangement as seen in Fig. 10c. Therefore, adding even one more degree of complexity by introducing a second set of particles opens up a whole new range of structural configurations tunable by multi-field combinations.

\section{Conclusions}
In summary, we have demonstrated multi-field, multi-stage assembly of complex, colloidal superstructures. By tuning the field parameters, we accessed a regime where magnetic and electric dipolar interactions and electrohydrodynamic (EHD) forces are comparable, leading to a rich variety of structural configurations. We showed that under parallel electric and magnetic fields, either a cooperation or a competition is observed depending on the frequency of the electric field. While cooperation among the electric and magnetic forces leads to the formation of large vertical stacks along the field, competition produces smaller stacks embedded among a quasi-planar cluster of particles. Under crossed electric and magnetic fields, a hierarchical structure is produced where each field contributes, leading to multi-scale structural reorganisation. While the in-plane magnetic field leads to the formation of chains, the electrohydrodynamic flows generated by the AC electric field at lower frequency values lead to the formation of quasi-planar non close-packed structures. The result is the formation of domains or clustered regions, which are individually composed of chain segments. The geometry of these hierarchical structures can be reversibly tuned by changing the magnetic field strength, frequency and voltage of the applied electric field. A higher magnetic field and a higher electric field frequency with a lower value of the electric field amplitude favors the formation of longer chains, while the opposite leads to the formation of clusters of individual particles. A detailed quantitative analysis of the structures formed under different values of the field parameters was done by taking the Fast Fourier Transforms of the obtained images, and subsequently calculating the angular and radial intensity profiles. This quantitative approach confirmed and extended the qualitative observations from microscopy, providing a rigorous framework for correlating field conditions with emergent structures. We also observed that the sequence of application of the two fields matters in determining the final structural configurations, with the field applied first having a dominating contribution. Lastly, we added one more level of complexity to our system by producing a binary mixture of magnetic and non-magnetic particles. This showed the formation of `poles' formed of magnetic particles connected by radial bridges of non-magnetic and magnetic particles, which could also be reconfigured under similar field parameter values. Our work demonstrates the ability to program multi-scale self-assembly through the balance of dipolar and EHD forces. Beyond advancing the fundamental understanding of field-directed assembly, this opens up opportunities for designing tunable photonic crystals, adaptive and switchable soft materials, and modular microswimmers with potential applications in targeted drug delivery and environmental remediation.

\section{Experimental Section}
\subsection{Sample preparation}
1 $\mu$m superparamagnetic polystyrene particles (Promag) were purchased from Bangs Laboratories. The zeta potential of the particles in deionized water was measured to be --40 mV. For non-magnetic particles used in the binary mixture experiments, 1$\mu$m polystyrene particles were also purchased from Bangs Laboratories. For field-based experiments, a suspension of the particles dispersed in deionized water was prepared with a solid content of 0.01 wt\%. A small concentration (4.33 mM) of a surfactant- sodium dodecyl sulphate (SDS) was added to ensure a smooth transfer to the capillary chamber. The capillary chamber was fabricated using two Indium Tin Oxide (ITO) coated glass cover-slips (thickness \#1). To prevent non-specific interactions between the particles and the substrate, the ITO-coated glass cover-slips were cleaned by sequential ultrasonic treatment in isopropyl alcohol for 15 minutes, acetone for 15 minutes, and again in isopropyl alcohol for 15 minutes. The cover-slips were then rinsed with deionized water and dried in an oven at 80°C. Two cleaned ITO-coated cover-slips were sandwiched and placed in a slightly displaced manner (see Fig. 1a) with a 150 $\mu$m plastic spacer (electrode separation) separating them. The ITO cover-slips function as electrodes, and placing them in a displaced manner ensures that electrical contacts can be smoothly made to these electrodes. The electrical contacts were made with copper tape and silver paste, as shown in Fig. 1a. The colloidal dispersion was then carefully pipetted into the capillary, and all ends were sealed using a UV-curable adhesive, ensuring that the particle dispersion remains contained within the capillary chamber. A sinusoidal AC electric field was applied using a METRAVI DDS-1010 function generator. A magnetic field was generated by a Helmholtz coil, with the sample positioned at the coil assembly's center to enable in situ magnetic field application and manipulation during observation. The magnetic field could be applied either in the x-y plane or along the z direction by reorienting the Helmholtz coils. The value of the magnetic field at the sample was measured using a Hall probe and a Gaussmeter. The experimental setup was observed under an upright optical microscope, utilizing a 100x objective lens in bright-field mode. Videos were recorded using a high-speed camera.

\subsection{Analysis}
For fields in cooperation and competition (sec 2.2 and Fig. 2) where both electric and magnetic fields were parallel to each other, bright objects in the focal plane were marked with red annotations and darker objects that were out of focus were marked with blue. In vertical stacks, the particles go out of focus, and hence, the stacks were marked with blue. In Fig. 2, the total number of objects counted is therefore a sum of the red and blue annotations. In Fig. 2a-c, the cooperative stacking from both the fields leads to every particle in the field of view joining a stack. Hence, the number of particles per stack ($N_s$) is calculated by dividing the total number of objects in Fig. 2a (no field condition) by the total number of stacks in Fig. 2c. Note that this is an approximate number. In the case of competitive interactions, as in Fig. 2f-h, the number of particles in the stack ($N_s$) is calculated as $$N_s = \frac{(\text{Total no. of objects in Fig. 2f}- \text{No. of red objects in Fig. 2h})}{\text{(No. of blue objects in Fig. 2h)}}.$$ This therefore excludes the particles (in red) that stay in the focal plane confined by the EHD forces from being counted in the calculation of $N_s$. Here, we assume that every blue object in Fig. 2h is a stack. This is a reasonable assumption since once the E field is turned ON, particles are brought into a planar configuration and hence would appear in the focal plane. Therefore, any instance of a particle going out of the focal plane would be due to stacking from the B field.

Fast Fourier transforms (FFTs) of the particle images (size $N \times N$) were computed in Python, as shown in Fig. 4b and 5b. The anisotropy in the FFTs was quantified by extracting angular intensity profiles from the FFTs, where the intensity at a given angle was obtained by summing the pixel intensities along the corresponding radial direction. This was then normalized by dividing by the minimum intensity value for each FFT. From the FFTs, the radial intensity profile was also obtained by calculating the total intensity over a circumference at a given radius value. This was normalized with the area under the curve. Two distinct peaks were identified in the resulting plot (Fig. 7b). The corresponding characteristic length scales were calculated from the peak positions using the relation,
\begin{equation}
\text{Length scale} = \frac{N \times 0.12}{\text{position}},
\end{equation}
where $0.12 \,\mu\text{m/pixel}$ is the image calibration factor. The peak at the lower radial value in the FFT represents the characteristic length scale associated with the chain segments (chain length and inter-chain spacing), while the peak at the higher radial value represents the length scale associated with individual particle size. Each of these two length scales was then normalized by dividing by the largest measured value of the corresponding length, and the bubble plots were generated (Fig. 7c-f). The un-normalized length scale corresponding to peak 1 was used in Fig. 9e and was referred to as the characteristic length.

\section*{Author contributions}
IB and SR fabricated the sample chambers and did the field-based experiments. IB performed the data analysis. IB, SR, and IC contributed to the planning of the experiments and writing of the paper.

\section*{Conflicts of interest}
There are no conflicts to declare.

\section*{Data availability}
Most of the data used in this work are either in the main paper or the ESI. Additional data will be available from the authors upon request as most of these are video data and are too bulky to store in a repository.

\section*{Acknowledgements}
We acknowledge the Start-up grant (SRG/2021/001696) of the Science and Engineering Research Board (SERB) (at present the Anusandhan National Research Foundation (ANRF)), Government of India and the additional competitive research grant (GOA/ACG/2021-2022/Nov/03) from BITS-Pilani for this work. We also acknowledge the central sophisticated instrumentation facility (CSIF) at BITS-Goa for the zeta-potential measurements.
\clearpage
%%%END OF MAIN TEXT%%%

%The \balance command can be used to balance the columns on the final page if desired. It should be placed anywhere within the first column of the last page.

\balance

%If notes are included in your references you can change the title from 'References' to 'Notes and references' using the following command:
\renewcommand\refname{References}

%%%REFERENCES%%%
\bibliography{rsc} %You need to replace "rsc" on this line with the name of your .bib file
\bibliographystyle{rsc} %the RSC's .bst file
\end{document}